%% file: main.tex
\documentclass{article}

\usepackage[preprint]{main}

\usepackage[utf8]{inputenc} 
\usepackage[T1]{fontenc}    
\usepackage{booktabs}       
\usepackage{amsfonts}       
\usepackage{microtype}      
\usepackage{xcolor}         
\usepackage[most]{tcolorbox}
\usepackage{caption}
\usepackage{adjustbox}
\usepackage{enumitem}
\usepackage{threeparttable}
\usepackage[ruled,linesnumbered]{algorithm2e}
\usepackage{wrapfig}
\usepackage{multirow}
\usepackage{subcaption}
\usepackage[normalem]{ulem}
\usepackage{hyperref}       

\useunder{\uline}{\ul}{}

\newcommand{\model}{{GeoSEE}} 

\definecolor{deeppink}{rgb}{1.0, 0.08, 0.58}

\newcommand{\cutparagraphup}{\vspace*{-0.1in}}

\title{\model{}: Regional Socio-Economic Estimation \linebreak With a Large Language Model} 

\author{%
Sungwon Han$^1$\thanks{Equal contribution.} \quad Donghyun Ahn$^{1*}$ \quad Seungeon Lee$^1$ \quad Minhyuk Song$^1$ \quad Sungwon Park$^1$ \\
\textbf{Sangyoon Park}$^2$ \quad \textbf{Jihee Kim}$^3$ \quad \textbf{Meeyoung Cha}$^{4,1}$ \\
$^1$School of Computing, KAIST \quad $^2$Division of Social Science, HKUST \\ 
$^3$College of Business, KAIST \quad $^4$ MPI for Security and Privacy \\
\texttt{\{lion4151, segaukwa, archon159, smh0706, psw0416\}@kaist.ac.kr}\\
\texttt{sangyoon@ust.hk, jiheekim@kaist.ac.kr, mia.cha@mpi-sp.org}
}

\begin{document}

\maketitle

\input{0_abstract}
\input{1_intro}

\input{2_method}

\input{3_experiments}

\input{4_related_work}
\input{5_conclusion}

\bibliographystyle{unsrtnat}
\bibliography{main}

\newpage
\appendix
\input{7_appendix}


\end{document}

%% file: 0_abstract.tex
\begin{abstract}
Moving beyond traditional surveys, combining heterogeneous data sources with AI-driven inference models brings new opportunities to measure socio-economic conditions, such as poverty and population, over expansive geographic areas.
The current research presents \model{}, a method that can estimate various socio-economic indicators using a unified pipeline powered by a large language model (LLM).
Presented with a diverse set of information modules, including those pre-constructed from satellite imagery, \model{} selects which modules to use in estimation, for each indicator and country.
This selection is guided by the LLM's prior socio-geographic knowledge, which functions similarly to the insights of a domain expert. The system then computes target indicators 
via in-context learning after aggregating results from selected modules in the format of natural language-based texts.
Comprehensive evaluation across countries at various stages of development reveals that our method outperforms other predictive models in both unsupervised and low-shot contexts. This reliable performance under data-scarce setting in under-developed or developing countries, combined with its cost-effectiveness, underscores its potential to continuously support and monitor the progress of Sustainable Development Goals, such as poverty alleviation and equitable growth, on a global scale. \looseness=-1
\end{abstract}
\vspace{-2mm}

%% file: 1_intro.tex
\section{Introduction}
Measuring socio-economic conditions at the subnational level is crucial for informed and data-driven decision-making in policy and business. This detailed assessment at a localized scale enables effective resource allocation and ultimately advances regional development. However, traditional surveys face significant challenges, including high costs, logistical complexities, and susceptibility to disruptions from natural disasters or conflicts, which impedes access to affected areas. In response to these challenges, the research community has begun to explore alternative data sources to supplement traditional data collection.
Examples include publicly available datasets such as Wikipedia text~\cite{sheehan2019predicting}, street view images~\cite{park2022deviancenet}, mobile phone adoption patterns~\cite{vscepanovic2015mobile}, and high-resolution satellite imagery~\cite{ahn2023human,albert2017using,han2020lightweight}. \looseness=-1

Current models that tap into these alternative data sources typically focus on predicting a single socio-economic indicator, such as population density, gross domestic product, or Gini coefficient, employing a limited number of data types~\cite{indaco2020twitter,jean2016combining,park2021improving}. Creating a universally applicable model that functions across multiple countries and indicators, while fully capitalizing on a diverse set of non-traditional data sources, is particularly challenging. One primary reason is that different regions and indicators show considerable variability in data availability and socio-economic characteristics. 
Additionally, each data source requires tailored methodologies to ensure accurate predictions, necessitating in-depth, specialized knowledge and substantial resources~\cite{head2017can}. This intensive need for expertise restricts the scalability and multimodality of these models. This issue is particularly pronounced in developing countries, where limited survey resources and a lack of comprehensive data coverage complicate the selection of suitable data sources and the development of reliable models for accurate predictions~\cite{ball2017comprehensive}.


We introduce \model{}, a universally applicable method that can estimate a diverse set of socio-economic indicators using a unified pipeline powered by a large language model (LLM).
The foundation of our approach is the concept of ``feature selection'' from multiple data sources in estimating socio-economic indicators~\cite{lewkowycz2022solving}.
Feature selection involves inferring associations between the input data and target labels.
This can be done using either a data-driven approach, which requires a sufficient amount of ground-truth labels, or an approach aided by domain experts. 
LLMs, with their vast repository of textual knowledge and reasoning capabilities~\cite{anil2023palm,openai2023gpt4}, can act as domain experts by selecting pertinent features from heterogeneous data sources to predict socio-economic indicators.
The methodology requires only the descriptions of the target indicator and each feature in natural language as a prompt, making it applicable even in underdeveloped countries which typically lack accurate ground-truth labels. \looseness=-1

\model{} first defines a list of modules to obtain enriched information from multiple data sources. 
These modules encompass techniques for processing satellite images, such as image segmentation, as well as methods for gathering data on Points of Interest (POI) or aggregating details about adjacent locales. 
Next, this array of modules and their respective descriptions are fed into the LLM as a prompt. 
This setup allows to select the modules most informative for solving the target problem based on prior knowledge. 
Feature selection, along with the self-consistency technique~\cite{wang2022self}, ensures reliable module selection even in the absence of ground-truth labels.
The final step involves collating data from these chosen modules to create a descriptive text paragraph about the target region. The text is then used in in-context learning to compare paragraphs from different regions and compute scores. 
When selecting in-context samples, we proposed a selection strategy that informs the LLM of the score distribution across different regions while also including regions similar to the target region in the in-context demonstration.

The primary strength of this work is its scalable multimodality for adding new modules and its capacity to predict multiple socio-economic indicators with a unified pipeline. 
We conducted experiments in two data-scarce scenarios, including when ground-truth labels are missing or only partially available, which are common in developing countries. 
We evaluate our model using various socio-economic indicators, including, but not limited to, population, education attainment, and labor force participation, across multiple countries at different stages of development. The results verify that our method generates predictions that align well with ground-truth labels, demonstrating broad applicability for monitoring the progress of Sustainable Development Goals (e.g., poverty reduction, equitable growth, urban green space) at a planetary-scale. \looseness=-1

%% file: 2_method.tex
\section{Methodology}
\subsection{Problem Statement and Overview}
\paragraph{Problem definition.} 
\model{} predicts regional socio-economic indicators even with scarce ground-truth data.
Consider a dataset $\mathcal{D}$ on $N$ regions of arbitrary shape and size (i.e., $\mathcal{D} = \{\mathbf{d}_i\}_{i=1}^N$) that encompass a substantial territory of a country. Subnational administrative units, such as districts, counties, and provinces are examples of the regions we consider. 
Then the main objective, a task of the model, is to estimate a socio-economic label $y_i$ for each region $\mathbf{d}_i$ in $\mathcal{D}$. 
We consider two scenarios: first is an unsupervised setting with no ground-truth labels (i.e., the set of unlabeled regions $\mathcal{D}_{ul} = \mathcal{D} = \{\mathbf{d}_i\}_{i=1}^N$); second is a $k$-shot setting with text description over a small number of $k$ regions (i.e., the set of region-label pairs $\mathcal{D}_{l} = \{(\mathbf{d}_i, y_i)\}_{i=1}^{k} \text{ and } \mathcal{D}_{ul} = \{\mathbf{d}_i\}_{i=k+1}^{N}$).

The model runs in two steps. 
In Step 1, within a list of information modules, the model selects a subset of modules to use, drawing on the LLM's existing knowledge base, for the given regions and a specific indicator to be assessed.
The selected modules are then applied to the target geographic region to extract task-specific information, which is then converted into text format using a predefined template (Section~\ref{sec:2.2}). 
After extracting text descriptions for each region in the dataset, in Step 2, the model leverages in-context learning by providing generated sample paragraphs of a few other regions as well as its own (Section~\ref{sec:2.3}). 
These regions are selected by our strategy designed to provide both detailed comparisons of similar regions and broader insights from the overall distribution of labels, while keeping the input text within the prompt limit. \looseness=-1


\subsection{Step 1: Task-Relevant Information Extraction via Module Selection}
\label{sec:2.2}
\paragraph{Module list.}
\model{} employs a range of internal information modules to compute socio-economic labels. Its flexible modular design allows for the easy addition of new data sources and functions. To optimize the processing of a diverse set of data sources and types, the model selects a subset of modules based on the LLM's prior knowledge and extracts only the information pertinent to the task. 

Our modules are engineered to gather all accessible public data about the given regions that can be relevant to socio-economic indicators. 
For instance, metrics such as nightlight intensity or overall luminosity captured in nighttime satellite images are indicative of economic activity levels.
Satellite images can also reveal land utilization patterns, called `landcover.' 
Points of Interest (POI) data may provide insights into a region's proximity to essential infrastructure, including airports and ports.
Moreover, leveraging geospatial metrics from adjacent areas can help estimate a region's socio-economic indicators~\cite{marshal1890, durantonetal2004}.
This set of information can be obtained directly from external databases like Nature Earth~\cite{kelso2010introducing} or deduced indirectly through analysis of satellite imagery~\cite{ahn2023human,huang2023learning}.
The complete list of modules used is as follows (see further details in the Appendix~\ref{sec:implementation_details}):
\begin{itemize}[leftmargin=3mm]
\item \texttt{get\_address}: Retrieves the address of a given region.
\item \texttt{get\_area}: Retrieves the area size of a given region.
\item \texttt{get\_night\_light}: Retrieves the nightlight intensity of a given region.
\item \texttt{count\_area}: Includes a set of modules that count the number of pixels that cover each of the target landcover classes (e.g., `road', `agricultural') and return the ratio of this count to the total number of pixels in the region's total image set.
\item \texttt{get\_distance\_to\_nearest\_target}: Includes a set of modules that measure the distance from a given region to each of the target class entities (e.g., `airport', `port').
\item \texttt{get\_aggregate\_neighbor\_info}: Includes a set of modules that retrieve information about neighboring regions using the functions defined above.
\end{itemize}

\begin{wrapfigure}{r}{0.4\textwidth}{
\vspace{-5mm}
\begin{tcolorbox}[enhanced,attach boxed title to top center={yshift=-1mm,yshifttext=-1mm},
  colback=black!0!white, colframe=black!20!white, colbacktitle=black!10!white, coltitle=blue!20!black ]
\footnotesize{
\texttt{Given a modular set, determine the sequence of modules that can be executed with inputs to solve the question given, following the format below. }\\
 \\
\texttt{Format for response:} \\
\texttt{1. MODULE 1} \\
\texttt{2. MODULE 2} \\
... \\
 \\
\texttt{The modules are defined as follows:} \\
$<$\texttt{Module Description}$>$ \\
 \\
\texttt{Question:} \\
\texttt{~~~$<$Task Description$>$} \\
\texttt{Input:}  \\
\texttt{- Location of the region - [Loc]} \\
 \\
\texttt{Answer:} }
\end{tcolorbox}
\captionof{figure}{Prompt for module selection in \model{}. An example of a full prompt is shown in Appendix~\ref{sec:appendix-A}. \looseness=-1}
\label{fig:prompt_for_module_selection}
\vspace{-2mm}
}
\end{wrapfigure}

\cutparagraphup
\paragraph{Module selection.} 
For each estimation task of an indicator in a target country, \model{} selects pertinent modules with the prompt. 
This prompt is an instruction for LLM to generate a response of module selection results, consisting of a module description and a target task description, as in Figure~\ref{fig:prompt_for_module_selection}.
\textit{Module description} includes functional specifications along with the input parameters it requires, for example: ``\texttt{get\_area(Loc): Get the area size of a given location's region}.'' 
\textit{Task description} states the indicator and the target country, for example, ``\texttt{what information is appropriate to infer Vietnam's regional GDP?}''  More prompt examples are given in the Appendix. \looseness=-1

The model takes this prompt as input and proposes potential module candidates for the given task. 
LLMs can inherently generate diverse logical pathways, each comprising a unique module combination.
For reliable module selection, our method involves multiple queries--specifically, ten iterations—to identify modules that are recommended at least five times. 
This approach aligns with the concept of self-consistency~\cite{wang2022self}, which posits that frequently used outcomes are more likely to be correct. \looseness=-1

Selected modules are then applied to each region in a target country, and the retrieved information is serialized into text using a predefined template. This process results in a comprehensive paragraph that represents the key features of the region:
\begin{align}
\text{Serialize}(f_1, \cdots, f_m, r_1, \cdots, r_m) = ``\ f_1\ \text{is}\ r_1.\ \cdots \ f_m\ \text{is}\ r_m."
\end{align}
where $f_1, \cdots, f_m$ are brief descriptions of the selected modules, and $r_1, \cdots, r_m$ are the results obtained from each module.

\subsection{Step 2: Estimation via In-Context Learning}
\label{sec:2.3}

After receiving natural language-based paragraphs for each region, LLM estimates the region's target indicator via in-context learning. 
We improve accuracy by expanding LLM's inference context to neighboring regions: we add paragraphs and estimation results of other regions as example demonstrations to the prompt. This provides multiple points of comparison to the model, allowing regions to be scored comparatively.
However, in few-shot or unsupervised scenarios where ground-truth labels are scarce, the number of examples available for comparison in the demonstrations can be insufficient.
Our model addresses this by saving a region's LLM inference scores (i.e., estimations) as pseudo-labels.
These pseudo-labels can be added to the prompt for in-context learning as pseudo-example demonstrations, providing a sufficient number of regions for comparison.

Algorithm~\ref{algo:algorithm} describes how the model infers scores for the given indicator of regions in the target country, using in-context learning. \looseness=-1
In-context learning here operates as follows: We start with a zero-shot or in-context learning-based inference for the unsupervised or few-shot setting for the first random region $\mathbf{d}_{\text{init}}$ in the unlabeled dataset $\mathcal{D}_{ul}$. 
Multiple identical queries (e.g., 3 times) with nonzero temperature (e.g., 0.5) are used to improve the initial region's inference accuracy by averaging the estimated values, similar to a recent study~\cite{wang2022self}. 
This inferred score is stored in the pseudo-labeled dataset $\mathcal{D}_{pl}$ (see L4-7 in Algorithm~\ref{algo:algorithm}).
Next, regions with estimated scores from $\mathcal{D}_{pl}$ and all samples from $\mathcal{D}_{l}$ are added as in-context demonstrations. 
Since $\mathcal{D}_{l}$ is empty in unsupervised setting, samples are only taken from $\mathcal{D}_{pl}$.
Newly estimated regions are moved from the unlabeled dataset $\mathcal{D}_{ul}$ to the pseudo-labeled dataset $\mathcal{D}_{pl}$. The process is repeated until $\mathcal{D}_{ul}$ becomes empty. 
Estimated values in $\mathcal{D}_{pl}$ become the final predictions (see L9-13 in Algorithm~\ref{algo:algorithm}).

\input{algo}

\cutparagraphup
\paragraph{Selection strategy for in-context demonstrations.}
Due to the limit in prompt length, only a limited set of pseudo-labels $\mathcal{D}_{pl}$ can be used as in-context demonstrations. 
We introduce a strategy for selecting pertinent in-context examples that can most contribute to the accuracy of LLM's estimations (see L10 in Algorithm~\ref{algo:algorithm}).  
The criteria for our selection strategy are twofold:
\begin{enumerate}[leftmargin=*]
\item \textit{Select examples that inform the LLM of the current pseudo-labels'  score distribution.} 
This gives a coarse-grained indication of where the score might fall within the distribution and prevents the model from deviating far from the score range.

\item \textit{Select similar examples to the target region regarding task-relevant information.}
Regions with similar task-relevant information are likely to yield similar scores, providing a fine-grained hint for score estimation.
\end{enumerate}

To implement the first criterion, the model sorts regions in the pseudo-labeled dataset $\mathcal{D}_{pl}$ by estimated scores and selects $n_{\text{coarse}}$ regions based on their ($n_{\text{coarse}}-1$)-quantile distribution. 
By dividing the score distribution into equal parts and presenting these examples to the LLM, we approximate the score distribution.
For the second criterion, the model selects $n_{\text{fine}}$ regions from $\mathcal{D}_{pl}$ that have similar task-relevant information to the target region. 
To assess information similarity between regions, we use only numerical outputs from modules\footnote{The only non-numerical module is `\texttt{get\_address}(Loc)', which is unsuitable as an indicator of similarity.}, concatenating them into a vector for each region (see Eq.~\ref{eq:r_i}). 
Next, we measure similarity between region vectors using the negative Euclidean distance after normalizing the vectors across all regions (see Eq.~\ref{eq:sim_i_j}). 
\looseness=-1
\begin{align}
&\mathbf{r}^i = \text{Concat}(\{ r^i_j | r^i_j \in \mathcal{R} \text{ and } j \in [1 ..m]\}) \label{eq:r_i} \\
&\text{sim}(\textbf{r}^{i_1}, \textbf{r}^{i_2}) = -|| \mathbf{r}^{i_1} - \mathbf{r}^{i_2} ||_2^2, \label{eq:sim_i_j}
\end{align}
where $r^i_j$ represents the result produced by the $j^{th}$ selected module for region $i$, $m$ is the total number of selected modules, and $i_1, i_2$ are the indices of the two regions.
Finally, we added a total of $n_{\text{coarse}}+n_{\text{fine}}$ region candidates to the set of in-context demonstration regions $\mathcal{B}_{\text{in-context}}$. \looseness=-1

%% file: algo.tex
\setlength{\textfloatsep}{2pt}
\begin{algorithm}[t!]
\SetAlgoLined
\small{
\SetKwInOut{Input}{Input}
\SetKwInOut{Output}{Output}
\Input{Large language model $F$, unlabeled dataset $\mathcal{D}_{ul}$, \linebreak labeled dataset $\mathcal{D}_l$ (for few-shot setting), a set of results from selected modules $\mathcal{R}$, \linebreak hyper-parameters $n_{\text{coarse}}$, $n_{\text{fine}}$.}
\Output{Pseudo-labeled dataset $\mathcal{D}_{pl}$}
$\mathcal{D}_{pl} \gets \emptyset$ \\
\While{$\mathcal{D}_{ul} \neq \emptyset$}{
    \If{$\mathcal{D}_{pl} = \emptyset$}{
        $\mathbf{d}_{\text{init}} \gets \text{Sample}(\mathcal{D}_{ul}$, 1) \\
        $(\mathbf{d}_{\text{init}}, \hat{y}_i) \gets F( \text{target}=\mathbf{d}_{\text{init}}, \text{ in-context}=\mathcal{D}_l, \text{ modules}=\mathcal{R}, \text{ queries}=3)$ \\
        $\mathcal{D}_{pl} \gets \{(\mathbf{d}_{\text{init}}, \hat{y}_i)\}$  \\
        $\mathcal{D}_{ul} \gets \mathcal{D}_{ul} - \{\mathbf{d}_{\text{init}} \}$       
    }
    $\mathbf{d} \gets \text{Sample}(\mathcal{D}_{ul}$, 1) \\
    $\mathcal{B}_{\text{ in-context}} \gets \text{SampleSelection}(\mathcal{D}_{pl}, \mathcal{R}, \mathbf{d}, n_{\text{coarse}}, n_{\text{fine}})$ \\
    $(\mathbf{d}, \hat{y}) \gets F(\text{target}=\mathbf{d}, \text{in-context}=\mathcal{B}_{\text{in-context}} \cup \mathcal{D}_l, \text{ modules}=\mathcal{R}, \text{ queries}=1)$ \\
    $\mathcal{D}_{pl} \gets \mathcal{D}_{pl} \cup \{(\mathbf{d}, \hat{y}) \}$  \\
    $\mathcal{D}_{ul} \gets \mathcal{D}_{ul} - \{\mathbf{d} \}$        
}
}
\caption{Estimation for given regions via in-context learning}
\label{algo:algorithm}
\end{algorithm}
\setlength{\textfloatsep}{15pt}

%% file: 3_experiments.tex
\section{Evaluation}
\subsection{Data and Implementation}
Countries at various stages of development were considered: a developed country (South Korea), an emerging country (Vietnam), and two least developed countries (Malawi and Cambodia), along with their daytime/nighttime satellite imagery and the POI data.
The daytime imagery was pulled from WorldView-2 and GeoEye, encompassing 2,223,408 images taken between 2018 and 2022, each with a spatial resolution of 2.4 meters and 256x256 pixels.
Nighttime imagery was procured from the Earth Observation Group (EOG) at a spatial resolution of 500 meters~\cite{elvidge2021annual}, where the data snapshot from 2022 was used.
Five socio-economic indicators were collected to evaluate the model's performance: regional GDP (GRDP), population (POP), elderly population (ELP), highly educated population ratio (HER), and labor force participation rate (LPR).
The ground-truth data were derived from the official websites of each respective country's government, as described in  Appendix~\ref{sec:dataset_details}.

Our framework is built upon GPT-4. The default values for top-p and temperature were 1 and 0.5. 
For in-context demonstrations, both $n_{\text{coarse}}$ and $n_{\text{fine}}$ were set to 5 for the unsupervised setting and 3 for the few-shot setting. 
These hyper-parameters were set according to the budget and prompt limits. 
While a higher setting can provide more information for inference, it also introduces a trade-off with increased costs. 
For implementation details for each module used in \model{}, refer to Appendix~\ref{sec:implementation_details}.

\subsection{Performance Comparison}
We consider unsupervised (i.e., no labels) and few-shot (i.e., five ground-truth labels available at the region-level) settings.
We employ both Pearson ($\rho_p$) and Spearman ($\rho_s$) correlation coefficients to measure agreement between our predictions and the ground-truth.
In the unsupervised context, we present the absolute values of these correlations ($|\rho_p|, |\rho_s|$) for fair comparison.
This is because, in the absence of labels, the relationship between the estimated scores and the ground-truth is unknown in several baselines.
To ensure robustness, we repeated the experiments three times using random seeds and divisions of labeled and unlabeled data.

We evaluated against four baselines for the unsupervised setting: 
(1) \textsf{Nightlight}~\cite{bagan2015analysis}: Uses scores based on the light intensity from nighttime satellite imagery of the region; 
(2) \textsf{SiScore}~\cite{han2020learning}: A human-in-the-loop model that trains a daytime satellite image-based scorer using coarse-grained human annotations; 
(3) \textsf{UrbanScore}~\cite{park2022learning}: Annotates a subset of daytime satellite images as urban, rural, or uninhabited and then trains a scorer using ordinal regression; 
(4) \textsf{GPT-4-Wiki}: 
Inspired by prior research~\cite{sheehan2019predicting}, this model extracts relevant paragraphs from  Wikipedia entries on the target region for zero-shot inference using the GPT-4 model.

We evaluated against seven baselines for the few-shot setting:
(1) \textsf{Nightlight}~\cite{bagan2015analysis}: Similar to the unsupervised Nightlight model but utilizes ground-truth labels to fit a linear model; 
(2) \textsf{SimpleCNN}: Fits an ImageNet-pretrained convolutional neural network (CNN) model using satellite imagery and few-shot labels to serve as a scorer; 
(3) \textsf{READ}~\cite{han2020lightweight}: Uses a CNN trained on a human-annotated dataset to summarize embeddings of satellite images within a region into a fixed-sized vector, then trains a regressor on this vector; 
(4) \textsf{Tile2Vec}~\cite{jean2019tile2vec}: This unsupervised representation learning model on satellite imagery is fitted on given few-shot region images and labels to serve as a scorer; 
(5) \textsf{SimCLR}~\cite{chen2020simple}: Similar to Tile2Vec, it performs unsupervised contrastive learning and then trains a regressor on the embeddings; 
(6) \textsf{GeoLLM}~\cite{manvi2023geollm}: Generates prompts using addresses and nearby locations within the region and fine-tunes a GPT-3.5 model on the training set; 
(7) \textsf{GPT-4-Wiki}: Uses the same Wikipedia paragraphs as in the unsupervised GPT-4-Wiki setting but includes training samples in the in-context demonstration.
Details on the implementation of each baseline follow the original work's setting and can be found in Appendix~\ref{sec:baseline_details}.

\input{tables/main_unsupervised}

\input{tables/main_fewshot_5}

\cutparagraphup
\paragraph{Results.}
Tables~\ref{table:main_unsupervised} and~\ref{table:main_fewshot_5} show performance comparisons for the unsupervised and 5-shot settings.
In the unsupervised setting, despite some fluctuations across metrics, our model has the highest average performance of 0.53 across all countries; the next best method is Nightlight, with 0.49.
Using textual data alone, as in GPT-4-Wiki, results in the lowest performance of 0.40.
In the few-shot setting, our model exhibits a substantial increase in Pearson correlation and consistently ranks among the top-2 results.
This remarkable rise in performance underscores the efficacy of incorporating label distribution data via in-context learning.
Full results with standard deviations, new indicators (e.g., gross regional domestic product or GRDP), and evaluation metrics (e.g., Spearman) are reported in Appendix~\ref{sec:full_results}. \looseness=-1

\input{tables/main_ablation_fewshot_5}
\subsection{Ablation Study}
To test the impact of each component, we removed or modified the following components one at a time:
(1) \textsf{Regression with all modules}: collecting numerical results from all modules in the module list, except the address, to perform linear regression on few-shot labels; 
(2) \textsf{Regression with selected modules}: collecting numerical results from selected modules via our prompt, except the address, to perform linear regression on few-shot labels; 
(3) \textsf{GPT-4 Address only}: conducting LLM-based inference using only address information, without modules; 
(4) \textsf{GPT-4 Address with neighbor only}: performing LLM-based inference using our model with prompts used in GeoLLM~\cite{manvi2023geollm}; 
(5) \textsf{Without coarse-grained selection}: using only the fine-grained selection criterion for in-context learning in \S\ref{sec:2.3};
(6) \textsf{Without fine-grained selection}:  using only the coarse-grained selection criterion;
(7) \textsf{Random selection}: randomly selecting in-context samples without our selection strategy. \looseness=-1

Table~\ref{tab:ablation} shows that any component-wise alteration or exclusion results in decreased performance. 
We make several observations. 
First, models that do not use LLM for inference (Ablations 1 and 2) perform significantly worse than the full model. 
This finding suggests that LLM's reasoning ability likely contributes to generalization in a low-shot setting, thus enhancing inference accuracy.
Second, between Ablations 1 and 2, the latter used less information but still performed comparably (and sometimes better) than Ablation 1. This trend demonstrates that our selection strategy can effectively identify the necessary information.
Third, the results of Ablations 3 and 4 demonstrate that utilizing information beyond the address with our modules leads to improved performance.
Finally, the results of Ablations from 5 to 7 collectively demonstrate that our in-context sample selection strategy enhances performance by enabling meaningful comparisons across regions.
We report ablation results for the unsupervised setting in Appendix~\ref{sec:appendix_ablation}. \looseness=-1

\section{Discussion}
So far we investigated the use of LLM to infer socio-economic indicators from geospatial data and reported promising results. 
Here, we discuss the operational mechanics of the model, its practical applications, and the various scenarios it addresses.

\paragraph{Is the LLM repeating memorized information?}
To test if learning occurs beyond the prior knowledge of the language model, we considered two variant models in a 5-shot setting. 
The first variant, \textsf{GPT-4 Address only}, corresponds to the third model in the ablation.
The other variation, \textsf{\model{} with permutations}, introduces noise by permuting module outcomes across regions to randomize the data.
These variants will reveal whether the specifically `curated' information from modules for each region was informative as opposed to using only the address or a random set of module outcomes.   \looseness=-1

\begin{figure}
    \centering
    \begin{subfigure}{.58\textwidth}
        \centering
        \raisebox{0mm}{
        \includegraphics[width=0.95\textwidth]{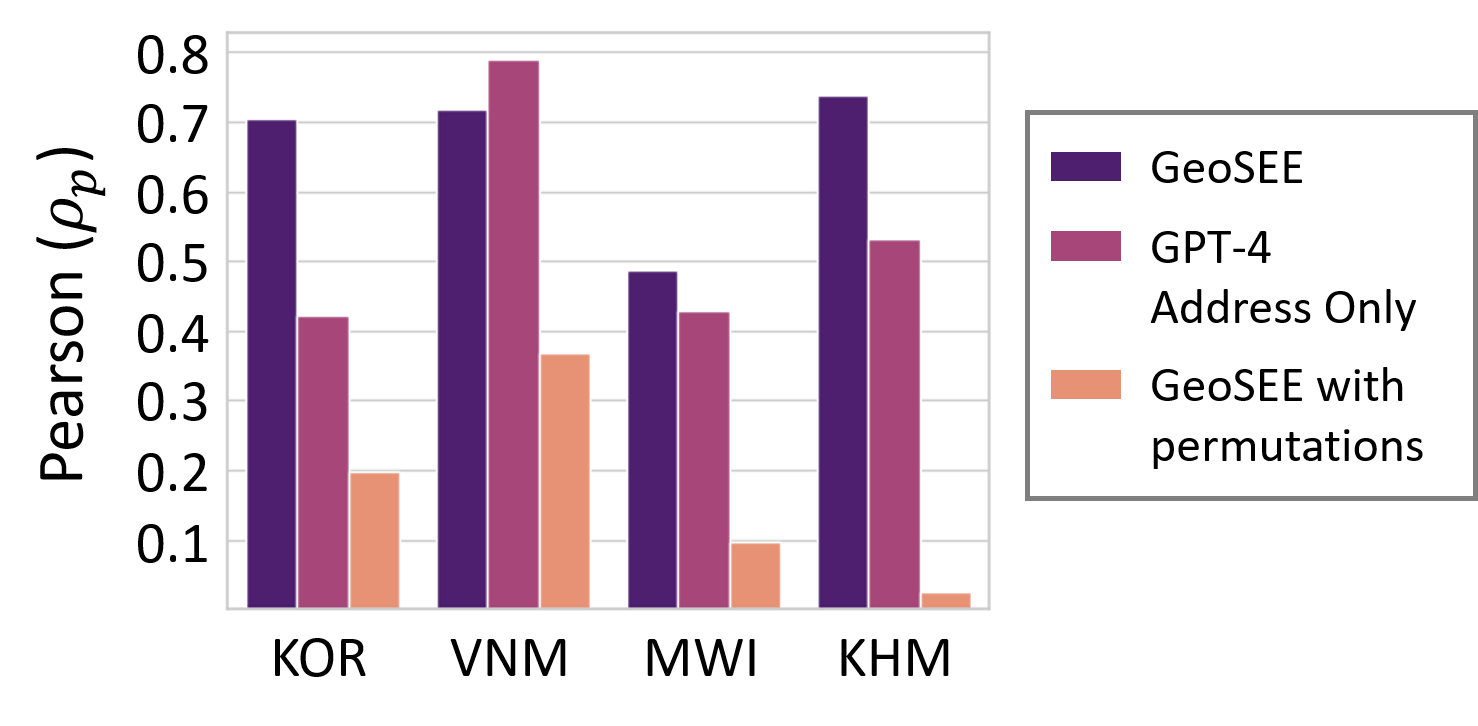}}
        \vspace*{-1mm}
        \caption{Effect of module outputs for inference quality}
        \label{fig:discussion1}
    \end{subfigure}
    \begin{subfigure}{.4\textwidth}
        \centering
        \raisebox{4mm}{
        \includegraphics[width=0.8\textwidth]{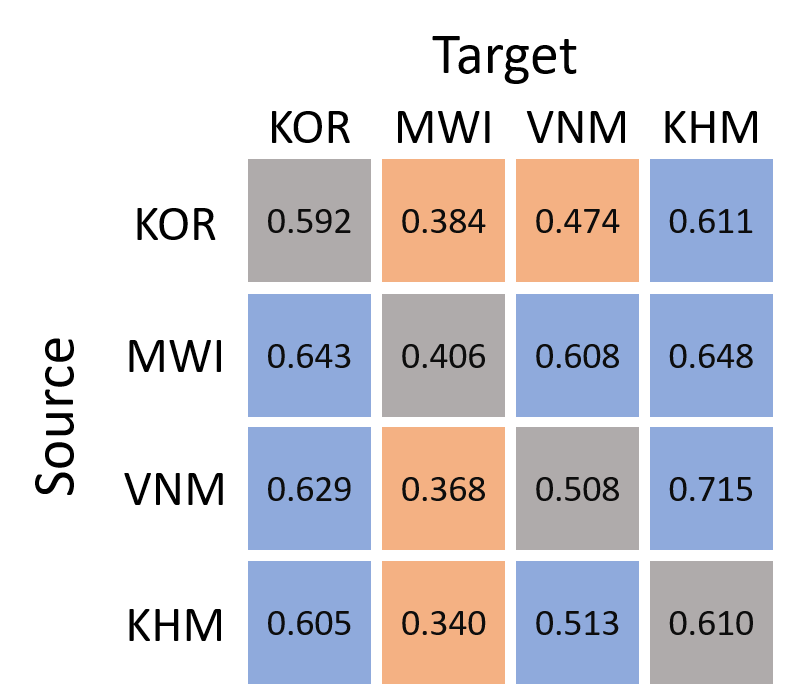}}
        \vspace*{-1mm}
        \caption{Transferability analysis}
        \label{fig:discussion2}
    \end{subfigure}
    \caption{ (a) Averaged Pearson correlation over four indicators (POP, ELP, HER, LPR) for each country in a 5-shot setting shows that our model's module improves LLM inference beyond the prior knowledge. (b) Averaged Pearson correlation transferring from a country (i.e., source country) to another country (i.e., target country). Rows represent the source country, and columns represent the target country; the diagonal line indicates evaluations in an unsupervised setting without transfer. \looseness=-1}
    \label{fig:discussion}
\end{figure}

Figure~\ref{fig:discussion1} shows compelling evidence for model learning.
Using the address-only model, we confirm that the LLM's prior knowledge alone can contribute to a meaningful Pearson correlation (above 0.4) for all investigated countries.
Interestingly, memorized knowledge performs best in Vietnam.
When the proposed modules are added, performance improves substantially even for Malawi and Cambodia, where the LLM has less prior knowledge about the target region.
Furthermore, when module results are shuffled, performance suffers considerably compared to using only addresses, indicating that module results are informative for the inference.

\cutparagraphup
\paragraph{Are module outcomes transferable across countries?}
This question asks the potential to apply module outputs and ground-truth data from one country (i.e., source) to another (i.e., target). We designed an experiment to test this idea by giving five in-context learning demonstrations selected from a source country to predict the indicators for the target country.
The LLM prompt was then updated to include the corresponding module output-induced paragraphs of selected regions. The criteria for selecting regions remained consistent with the original model.

Figure~\ref{fig:discussion2} depicts the transferability potential based on average Pearson correlations.
The analysis shows that, in most cases, the model's insights are transferable, as evidenced by the comparative improvements over the unsupervised version of \model{} that appear in the diagonal line.
Malawi is an exception, where transfer learning underperforms against the unsupervised scenario.
This divergence could be attributed to Malawi's unique African geographic landscape, which likely differs substantially from that of other Asian countries. 
We leave the task of designing a selection strategy tailored for efficient transfer learning pairs as future work.

\begin{figure}[!ht]
    \centering
    \includegraphics[height=5cm]{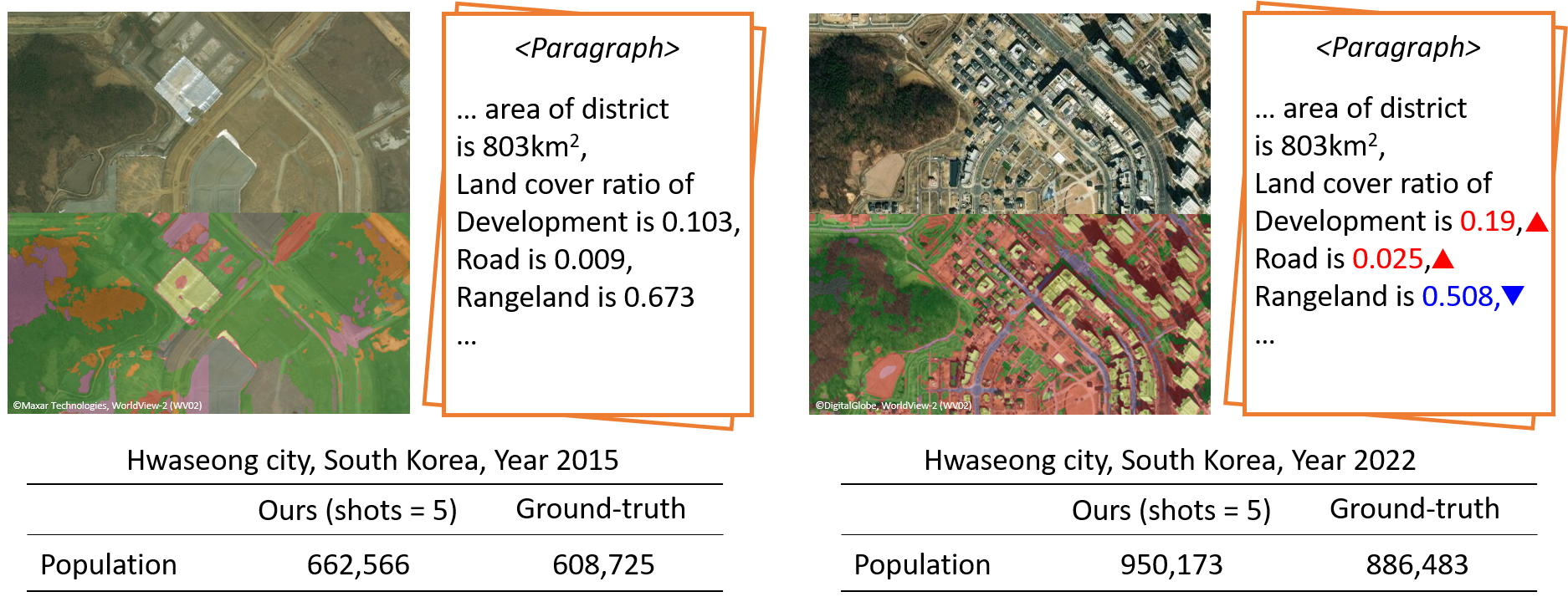}
    \caption{Qualitative analysis on predicting changes between two timestamps, providing example satellite images, segmentation maps and paragraphs. The analysis is done over the Hwaseong City area in South Korea. Differences captured by the modules constructed from satellite imagery lead to different estimation results, which show positive growth in the area, consistent with ground-truths. {\footnotesize (Note that colored texts and triangle symbols are illustrations here only, and not given to the LLM.)} \looseness=-1}
    \label{fig:discussion3}    
\end{figure}

\cutparagraphup
\paragraph{Can \model{} detect changes over time?}
Geospatial characteristics change over time, and capturing these changes is critical for many applications. 
We conducted a qualitative analysis to see whether our model can detect changes in socioeconomic conditions over time. We present a case study of Hwaseong City in South Korea, which had considerable growth between 2015 and 2022, as evidenced by changes in various indicators. 
We ran \model{} to predict the city's population in 2015 and 2022 each, using demonstration samples from five randomly selected regions (i.e., 5-shot) that excluded the target region. 
Figure~\ref{fig:discussion3} shows that the model's estimation results are consistent with the actual population growth trend in the area. While more robust analyses are needed to generalize this finding, this case study hints at the model's potential for tracking and analyzing changes over time. \looseness=-1

\cutparagraphup
\paragraph{How does \model{}'s module selection differ by country and indicator?} 
We report which information modules are selected by \model{}, depending on a country and indicator, in Table \ref{table:appendix_module_selection} of the Appendix~\ref{sec:appendix_module_selection}. Some notable observations from the table are: (1) module selection varies both by country and by indicator. For instance, in the estimation of population in South Korea, 12 modules are used as opposed to 8 modules in higher education. GRDP is a good example to show variations across countries: both its own and neighboring region's agricultural landcovers were chosen for Malawi, Vietnam, and Cambodia, but not for South Korea. This is an economically intuitive result considering the countries' development stages. (2) Some modules---address, area size, and nightlight--- are used in every case - for all the countries and indicators, whereas modules like the water landcover are never chosen. 
(3) We also spot some consistency in module selection across countries. Specifically, agricultural landcover is used to estimate (total) population and labor force participation rate in South Korea, Malawi, and Cambodia. Except for Malawi, agricultural land cover is not used to predict the elderly population. Further analysis in this vein can help us develop strategies to enhance the model's interpretability, shedding light on the underlying relationships between indicators and geospatial data.   \looseness=-1

%% file: tables/main_unsupervised.tex
\begin{table}[t!]
\centering
\setlength{\tabcolsep}{2pt}
\caption{Performance evaluation results based on Pearson correlation in unsupervised settings. Best results are marked in bold text, and our model's results are underlined when it is second-best.}
\label{table:main_unsupervised}
\renewcommand{\arraystretch}{1.4}
\resizebox{1\textwidth}{!}{
\begin{tabular}{@{}l|cccccccc|cccc|cccc|c@{}}
\toprule
\multirow{2}{*}{Method} & \multicolumn{4}{c|}{South Korea (KOR)} & \multicolumn{4}{c|}{Malawi (MWI)} & \multicolumn{4}{c|}{Vietnam (VNM)} & \multicolumn{4}{c|}{Cambodia (KHM)} & \multirow{2}{*}{Avg} \\ \cline{2-17}
 & POP & ELP & HER & \multicolumn{1}{c|}{LPR} & POP & ELP & HER & LPR & POP & ELP & HER & LPR & POP & ELP & HER & LPR &  \\ \hline
Nightlight & \textbf{0.70} & \textbf{0.63} & 0.53 & \multicolumn{1}{c|}{0.55} & \textbf{0.43} & 0.09 & \textbf{0.86} & 0.19 & 0.25 & 0.23 & \textbf{0.58} & 0.01 & 0.76 & 0.64 & \textbf{0.83} & 0.51 & 0.49 \\
SiScore & 0.46 & 0.49 & 0.64 & \multicolumn{1}{c|}{\textbf{0.70}} & 0.03 & 0.18 & 0.80 & 0.11 & 0.37 & 0.45 & 0.08 & 0.14 & 0.75 & \textbf{0.74} & 0.74 & 0.36 & 0.44 \\
UrbanScore & 0.38 & 0.42 & 0.61 & \multicolumn{1}{c|}{0.66} & 0.08 & 0.27 & 0.75 & 0.21 & 0.42 & 0.46 & 0.43 & 0.09 & 0.64 & 0.60 & 0.76 & 0.43 & 0.45 \\
GPT-4-Wiki & 0.54 & 0.31 & 0.56 & \multicolumn{1}{c|}{0.28} & 0.33 & \textbf{0.29} & 0.65 & \textbf{0.28} & 0.67 & 0.22 & 0.50 & 0.04 & 0.61 & 0.29 & 0.25 & \textbf{0.57} & 0.40 \\
\model{} & {\ul 0.63} & {\ul 0.47} & \textbf{0.65} & \multicolumn{1}{c|}{0.62} & {\ul 0.37} & 0.16 & {\ul 0.85} & {\ul 0.24} & \textbf{0.82} & \textbf{0.63} & 0.22 & \textbf{0.36} & \textbf{0.78} & {\ul 0.70} & 0.61 & 0.35 & \textbf{0.53} \\ \bottomrule
\end{tabular}
}
\end{table}

%% file: tables/main_fewshot_5.tex
\begin{table}[t!]
\centering
\setlength{\tabcolsep}{2pt}
\caption{Performance evaluation based on Pearson correlation in 5-shot settings. 
Some estimates of GPT-based models failed and reported the same values across all regions; we mark such a case `N/A'.}
\label{table:main_fewshot_5}
\renewcommand{\arraystretch}{1.4}
\resizebox{1\textwidth}{!}{
\begin{tabular}{@{}l|cccc|cccc|cccc|cccc|c@{}}
\toprule
\multirow{2}{*}{Method} & \multicolumn{4}{c|}{South Korea (KOR)} & \multicolumn{4}{c|}{Malawi (MWI)} & \multicolumn{4}{c|}{Vietnam (VNM)} & \multicolumn{4}{c|}{Cambodia (KHM)} & \multirow{2}{*}{Avg} \\ \cline{2-17}
 & POP & ELP & HER & \multicolumn{1}{c|}{LPR} & POP & ELP & HER & LPR & POP & ELP & HER & LPR & POP & ELP & HER & LPR &  \\ \hline
Nightlight & 0.55 & 0.48 & 0.53 & 0.55 & -0.02 & -0.14 & 0.30 & -0.08 & 0.05 & 0.02 & -0.22 & -0.01 & 0.68 & 0.59 & 0.70 & 0.44 & 0.28 \\
SimpleCNN & 0.24 & 0.30 & 0.49 & 0.24 & 0.37 & 0.49 & 0.05 & 0.23 & 0.19 & 0.41 & 0.03 & 0.41 & 0.04 & 0.16 & -0.30 & 0.16 & 0.27 \\
READ & 0.22 & 0.24 & 0.25 & 0.25 & -0.06 & -0.04 & 0.48 & 0.24 & 0.26 & 0.16 & -0.28 & 0.06 & 0.26 & 0.28 & 0.17 & -0.06 & 0.15 \\
Tile2Vec & 0.42 & 0.29 & 0.56 & 0.29 & 0.26 & 0.06 & 0.23 &\textbf{0.31} & 0.14 & 0.24 & 0.05 & \textbf{0.52} & -0.01 & -0.15 & 0.22 & 0.22 & 0.22 \\
SimCLR & 0.23 & 0.24 & 0.33 & 0.02 & 0.20 & 0.13 & 0.32 & 0.16 & 0.05 & 0.16 & 0.00 & -0.16 & 0.42 & 0.40 &  0.14 & 0.14 & 0.17 \\
GeoLLM & 0.63 & 0.01 & -0.02 & -0.10 & 0.40 & 0.49 & 0.00 & N/A & 0.84 & 0.80 & 0.56 & 0.01 & 0.33 & 0.12 & 0.48 & 0.11 & 0.31 \\
GPT-4-Wiki & 0.59 & 0.41 & 0.56 & -0.26 & 0.52 & \textbf{0.58} & 0.70 & 0.15 & 0.83 & 0.80 & 0.47 & 0.11 & 0.55 & 0.64 & 0.54 & -0.21 & 0.44 \\
\model{} & \textbf{0.67} & \textbf{0.75} & \textbf{0.71} & \textbf{0.69} & \textbf{0.75} & {\ul 0.51} & \textbf{0.94} & -0.26 & \textbf{0.99} & \textbf{0.90} & \textbf{0.65} & 0.32 & \textbf{0.73} & \textbf{0.79} & \textbf{0.83} & \textbf{0.60} & \textbf{0.66} \\
\bottomrule
\end{tabular}
}
\end{table}

%% file: tables/main_ablation_fewshot_5.tex

\begin{table}[t!]
\caption{Ablation results reporting Pearson correlations for each country and socio-economic indicators (i.e., POP, ELP, HER, LPR) after 
alteration or exclusion of components in \model{}. Statistics are averages based on the 5-shot setting. \looseness=-1} 
\centering
\begin{tabular}{@{}lcccc|c@{}}
\toprule
Model & KOR & MWI & VNM & KHM & Avg \\ \midrule
\model{}  & 0.704 & 0.486 & 0.717 & 0.739 & \textbf{0.662} \\
(Ablation 1) Regression with all modules & 0.429 & 0.309 & 0.254 & 0.462 & 0.363 \\
(Ablation 2) Regression with selected modules & 0.576 & 0.175 & 0.381 & 0.355 & 0.372 \\
(Ablation 3) GPT-4 Address only & 0.421 & 0.429 & 0.789 & 0.532 & 0.543 \\
(Ablation 4) GPT-4 Address with neighbor only & 0.501 & 0.340 & 0.823 & 0.261 & 0.481 \\
(Ablation 5) Without coarse-grained selection & 0.662 & 0.498 & 0.702 & 0.377 & 0.560 \\
(Ablation 6) Without fine-grained selection & 0.782 & 0.502 & 0.712 & 0.312 & 0.577 \\
(Ablation 7) Random selection & 0.654 & 0.539 & 0.716 & 0.391 & 0.575 \\ \bottomrule
\multicolumn{6}{l}{KOR: South Korea,~ MWI: Malawi, ~VNM: Vietnam, and ~KHM: Cambodia}
\end{tabular}
\label{tab:ablation}
\end{table}

%% file: 4_related_work.tex
\section{Related Work}
\paragraph{Proxy-based estimation.}
External data sources can supplement costly surveys, as in the case of satellite imagery~\cite{ahn2023human,albert2017using,han2020lightweight}, structured data like POIs~\cite{huang2023learning}, and social media posts~\cite{indaco2020twitter,paul2011you,signorini2011use}.  
For instance, light intensity of night-time imagery is known to strongly correlate with regional economic indicators~\cite{bagan2015analysis,ghosh2013using}. 
Recent methods use deep learning to predict indicators using daytime imagery and street views~\cite{jean2016combining,park2022learning,xi2022beyond}. 
POI data are also used to predict socio-economic factors at the regional level~\cite{huang2023learning}. 
Another study extracted textual embeddings from regional Wikipedia articles for prediction purposes~\cite{sheehan2019predicting}. 
These data act as proxies for survey data and are relatively easier to collect. 
Our research aligns with this approach, employing diverse datasets for computation.

\cutparagraphup
\paragraph{Limited labels.}
Training reliable estimators for socio-economic indicators often requires ground-truth data, which poses a challenge in developing countries.
Recent studies have proposed methods for using publicly available data in unsupervised, weakly-supervised, or semi-supervised manners.
One example is a human-in-the-loop structure for economic development~\cite{han2020learning}.
In another study~\cite{liu2021nightlight}, nightlight intensity was used as a pseudo-label to train a daytime satellite imagery-based encoder.
We propose a zero-shot and in-context learning model that can flexibly accommodate new data sources.

\cutparagraphup
\paragraph{LLM for geospatial data.}
LLMs can be helpful for many domain-specific tasks.
For example, GeoGPT pioneered GPT-based geospatial data aggregation, processing, and analysis~\cite{zhang2023geogpt}.
Other studies performed dementia forecasting using time series analysis and urban function classification using POI data~\cite{mai2023opportunities} and optimized models for geospatial data estimation~\cite{deng2023learning,manvi2023geollm}.
However, these studies rely on either verified data or the model's prior knowledge, thus have limited applicability to developing countries with scarce data (i.e., data gap) or less-known information that is also unlikely to be included in the LLM's prior knowledge (i.e., knowledge gap).
We overcome this limitation by incorporating diverse proxy data, utilizing the reasoning ability of the LLM as a feature selector. 

%% file: 5_conclusion.tex
\section{Conclusion}
We presented an LLM-based, universally applicable pipeline for estimating socio-economic indicators across diverse geographic settings.
\model{} is grounded in the principle of selecting key features from multiple data sources and available compute modules.
The LLM serves as a domain expert for this inference task, identifying relevant data points across the data sources based on its extensive prior knowledge and reasoning abilities. 
The simplicity of its structure, which only requires natural language descriptions of the desired indicator and features, makes the model adaptable and extensible, allowing computations on any geolocation, even in areas that have limited data.
\looseness=-1

\cutparagraphup
\paragraph{Limitations and broader impact.}
Several factors need to be considered going forward. 
Firstly, the experiments used a limited set of external data and modules. However, our model is easily expandable to accommodate more modules, so future studies can build on our findings using newly available data sources and modules. 
Secondly, our model is tested for a one-time snapshot for each country and indicator. While we demonstrated its potential for tracking temporal changes, further improvements are needed to enable reliable analyses at finer temporal intervals. 
This will be crucial for practical applications that require detailed time-scale insights.
The impact of our research is particularly significant in enhancing socio-economic analysis at the subnational level, offering critical measurements for informed policy and business decisions. Moreover, this work has the potential to facilitate the monitoring of sustainable development goals, especially in regions where resources for traditional data collection are limited.

%% file: 7_appendix.tex
\section*{Appendix}
\section{Example Prompts in \model{}}
\label{sec:appendix-A}

\subsection{Example Prompt for Module Selection}
We introduce an example prompt for module selection of \model{}. We assume the experimental setting for predicting population of regions in Malawi.

\begin{figure}[h!]
\begin{tcolorbox}[enhanced,attach boxed title to top center={yshift=-3mm,yshifttext=-1mm},
  colback=black!0!white, colframe=black!20!white, colbacktitle=black!10!white, coltitle=blue!20!black ]
Given a modular set, determine the sequence of modules that can be executed with inputs to solve the question given following the format below. \\
\\
Format for reseponse:\\
1. MODULE 1\\
2. MODULE 2\\
...\\
\\
The modules are defined as follows:\\
- count\_area(Loc, Class): Count the pixels of the given class in the location image. Class should be one of the element in ["bareland", "rangeland", "development", "road", "tree", "water", "agricultural", "building"].\\
\\
- get\_address(Loc): Get address of given location.\\
\\
- get\_area(Loc): Get area size of given location's region.\\
\\
- get\_night\_light(Loc): Get nightlight intensity of given location.\\
\\
- get\_distance\_to\_nearest\_target(Loc, Class): Get distance of given location to class. Class should be one of the element in ['airport', 'port']\\
\\
- get\_aggregate\_neighbor\_info(Loc, Func): Get neighbor regions' information of given location, using functions defined above. The format of Func would be the lambda function (i.e., lambda x: [function name](loc=x, ...)).\\
\\
Question: Which information is useful to infer the population of Malawi?\\
Input: \\
- Location of the region - [Loc]\\
\\
Answer: 
\end{tcolorbox}
\captionof{figure}{Example prompt for module selection of \model{} to predict population in Malawi.}
\label{fig:appendix-module_prompt} 
\end{figure}

\newpage
\subsection{Example Prompt for Inferring Scores of Region}
We show an example prompt for inferring population scores for a region in Malawi using our in-context learning approach. \looseness=-1

\begin{figure}[h!]
\begin{tcolorbox}[enhanced,attach boxed title to top center={yshift=-3mm,yshifttext=-1mm},
  colback=black!0!white, colframe=black!20!white, colbacktitle=black!10!white, coltitle=blue!20!black ]
full address of the given location is Kabwabwa, Lilongwe, Central Region, MWI. Land cover ratio of development is 0.229. Land cover ratio of building is 0.042. Land cover ratio of rangeland is 0.579. Land cover ratio of agricultural is 0.014. Sum of nightlight intensity is 8182.944. Average nightlight intensity is 3.824. area of region (km$^2$) of the given location is 402.647. Land cover ratio of development in the neighboring region(s) is 0.119. Land cover ratio of building in the neighboring region(s) is 0.003. Sum of nightlight intensity in the neighboring region(s) is 2651.147. Average nightlight intensity in the neighboring region(s) is 0.046. Land cover ratio of rangeland in the neighboring region(s) is 0.717. Land cover ratio of agricultural in the neighboring region(s) is 0.009. \\
Infer a the population from given location's description. Answer the numeric score only. \\
Answer: 1177613.0 \\
 \\
full address of the given location is Mzimba, Northern Region, MWI. Land cover ratio of development is 0.096. Land cover ratio of building is 0.003. Land cover ratio of rangeland is 0.706. Land cover ratio of agricultural is 0.013. Sum of nightlight intensity is 780.235. Average nightlight intensity is 0.02. area of region (km$^2$) of the given location is 10437.523. Land cover ratio of development in the neighboring region(s) is 0.045. Land cover ratio of building in the neighboring region(s) is 0.005. Sum of nightlight intensity in the neighboring region(s) is 3375.725. Average nightlight intensity in the neighboring region(s) is 0.031. Land cover ratio of rangeland in the neighboring region(s) is 0.427. Land cover ratio of agricultural in the neighboring region(s) is 0.006.\\
Infer a the population from given location's description. Answer the numeric score only.\\
Answer: 1030892.0\\
\\
... \\
\\
full address of the given location is Machinga, Southern Region, MWI. Land cover ratio of development is 0.087. Land cover ratio of building is 0.008. Land cover ratio of rangeland is 0.677. Land cover ratio of agricultural is 0.005. Sum of nightlight intensity is 347.965. Average nightlight intensity is 0.029. area of region (km$^2$) of the given location is 3910.786. Land cover ratio of development in the neighboring region(s) is 0.066. Land cover ratio of building in the neighboring region(s) is 0.004. Sum of nightlight intensity in the neighboring region(s) is 2267.136. Average nightlight intensity in the neighboring region(s) is 0.041. Land cover ratio of rangeland in the neighboring region(s) is 0.526. Land cover ratio of agricultural in the neighboring region(s) is 0.004.\\
Infer a the population from given location's description. Answer the numeric score only.\\
Answer: 885670.0\\
\\
full address of the given location is Dedza, Central Region, MWI. Land cover ratio of development is 0.114. Land cover ratio of building is 0.004. Land cover ratio of rangeland is 0.664. Land cover ratio of agricultural is 0.007. Sum of nightlight intensity is 456.095. Average nightlight intensity is 0.016. area of region (km$^2$) of the given location is 4008.478. Land cover ratio of development in the neighboring region(s) is 0.084. Land cover ratio of building in the neighboring region(s) is 0.004. Sum of nightlight intensity in the neighboring region(s) is 4359.157. Average nightlight intensity in the neighboring region(s) is 0.04. Land cover ratio of rangeland in the neighboring region(s) is 0.55. Land cover ratio of agricultural in the neighboring region(s) is 0.004.\\
Infer a the population from given location's description. Answer the numeric score only.\\
Answer: 
\end{tcolorbox}
\captionof{figure}{Example prompt for inferring population in Malawi. }
\label{fig:appendix-full_prompt} 
\end{figure}

\section{Implementation Details of \model{} }
\label{sec:implementation_details}
\subsection{Module Implementations}
Our model defines various modules to conduct estimation using either freely or academically available data sources. All administrative region and its boundary information are provided by ArcGIS REST API service.
The implementation details for each module are as follows:
\begin{itemize}[leftmargin=*]
\item \texttt{get\_address}: This function first retrieves the administrative region and its boundary for the specified location. It then conducts reverse geocoding on the centroid of the region to return the address.

\item \texttt{get\_area}: This function retrieves the administrative region and its boundary for the specified location, and computes the size of the boundary.

\item \texttt{get\_night\_light}: This refers to data from VIIRS nightlight imagery, which covers the entire globe. It crops the imagery to align with the administrative region boundary of the specified location and reports both the sum and average light intensity within the boundary.

\item \texttt{count\_area}: This module counts the number of pixels covering the target land-cover class using the deep learning segmentation model proposed in a previous study~\cite{buscombe2022reproducible}. The model performs segmentation on nine classes: bare land, rangeland, development, road, tree, water, agricultural, building, and nodata~\cite{xia2023openearthmap}. It then returns the ratio of this count to the total number of pixels in the image of the location.

\item \texttt{get\_distance\_to\_nearest\_target}: This function measures the distance from a specified location to a target class entity based on Natural Earth data~\cite{kelso2010introducing}.

\item \texttt{get\_aggregate\_neighbor\_info}: This function retrieves information about neighboring regions of a given location using the functions defined above. Two regions are considered `neighbors' if their boundaries share at least one point.
\end{itemize}

We use four NVIDIA GeForce RTX 3090 GPUs for running modules with parallelism. 
Typically, it takes less than 12 hours to generate the output for all modules in a single run.

\subsection{Inference Setting Details}
\model{} utilized GPT-4 as the LLM backbone. The top-p used for LLM inference was set to 1, which is the default setting of the API, and the temperature was set to 0.5. 
When selecting in-context demonstrations, both $n_{\text{coarse}}$ and $n_{\text{fine}}$ were set to 5 in unsupervised setting, and 3 for few-shot settings. \looseness=-1

\section{Baseline Details}
\label{sec:baseline_details}
We considered four baselines for the unsupervised setting and seven for the few-shot setting. 
Here below descriptions are the implementation details of each baseline.

\subsection{Baselines for Unsupervised Setting}
\begin{itemize}[leftmargin=*]
\item \textbf{Nightlight}: We investigate the direct correlation between the nightlight intensity from nighttime satellite imagery of the region and its socio-economic indicators. For target indicators expressed as ratios, we use the average nightlight intensity within the region. Conversely, for target indicators represented as estimated numbers, we use the sum of the nightlight intensity.

\item \textbf{SiScore}: This method assigns a score to each satellite image by maximizing the Spearman correlation between the estimated and actual ranks of the images. During the transfer learning phase of the clustering process, we utilized soft labels provided by four human annotators, who labeled 1,000 images across urban, rural, and uninhabited classes. A total of 21 clusters were used, with 10 clusters each for urban and rural classes and an additional cluster for the uninhabited class. Negative scores were clamped to zero in the final phase. Scores of the images within the region were averaged to represent the final score of the region.

\item \textbf{UrbanScore}: This method employs ordinal regression to score each satellite image, using a limited number of human labels. Four human annotators labeled 1,000 images among urban, rural, and uninhabited categories. Thresholds between urban-rural and rural-uninhabited classes were set at 0 and 10, respectively, and negative scores were clamped to zero in the final phase. Scores of the images within the region were averaged to represent the final score of the region.

\item \textbf{GPT-4-Wiki}: This method predicts socio-economic labels by utilizing paragraphs extracted from regional Wikipedia articles. Geolocated regional Wikipedia articles can be obtained by sending a query with specific latitude and longitude coordinates to the Wikipedia API. After collecting and summarizing Wikipedia articles at the sub-national level, the target-related Wikipedia paragraph is extracted using the prompt ``Extract the [target indicator] information from the following paragraph.'' For population indicators, if a numerical value is explicitly stated in the paragraph, the corresponding information is masked before the prediction is conducted.
\end{itemize}

\subsection{Baselines for Few-Shot Setting}
\begin{itemize}[leftmargin=*]

\item \textbf{Nightlight}: This method trains a linear regressor in a supervised manner using nightlight as a single feature. For target indicators expressed as ratios, we use the average nightlight intensity within the region. Conversely, for target indicators represented as estimated numbers, we use the sum of the nightlight intensity.

\item \textbf{SimpleCNN}: This method employs a convolutional neural network (CNN) to predict socio-economic labels. We fine-tuned an ImageNet-pretrained ResNet18 model with a linear regressor through supervised learning on images from a region. The labels for these images are assigned based on the socio-economic indicators of the region. If the target indicator is a ratio, the indicator's value for the region is directly used as the label. Conversely, if the target indicator is an estimated number, the model is trained to predict the logarithm of this number, divided by the number of images in the region.

\item \textbf{READ}: This method utilizes a small subset of human-labeled satellite images and a large number of unlabeled images to extract robust and lightweight image representations. Four human annotators labeled 1,000 images among urban, rural, and uninhabited categories. Following the original paper, uninhabited images were pruned from the dataset based on labels decided by majority votes from three annotators. Before obtaining the final region representation, the dimension of each image embedding was reduced to three using the PCA method.

\item \textbf{Tile2Vec}: Tile2Vec is an approach to unsupervised representation learning using satellite imagery. We utilized the pre-trained Tile2Vec model, which was made available on GitHub by the original authors. Similar to the SimpleCNN approach, we fine-tuned this model with a linear regressor through supervised learning on images from a region.

\item \textbf{SimCLR}: This method uses augmented images as contrastive samples to learn image representations. The set of augmentations followed the original SimCLR GitHub repository. Similar to the READ method, a pruned image dataset was used to train the model and extract image-level representations. Individual image representations were averaged to generate the final representation of the region, and an XGBoost regressor was used to predict the socio-economic labels.

\item \textbf{GeoLLM}: Following the method described in the original paper~\cite{manvi2023geollm}, prompts are constructed using the target region and neighboring locations. 
Subsequently, the model is trained on few-shot samples using the fine-tuning API of OpenAI's GPT-3.5-turbo model. 
Inference is then performed using the trained model. 
Training utilizes the API's default settings, with the number of epochs set to 20 and the learning rate multiplier set to 2.

\item \textbf{GPT-4-Wiki}: Similar to the unsupervised setting, information is sourced from Wikipedia to construct prompts. 
However, in the few-shot setting, unlike the zero-shot inference of the unsupervised setting, Wikipedia information and labels about regions within the training set are additionally used as in-context demonstrations.

\end{itemize}
\newpage
\section{Dataset Details}
\label{sec:dataset_details}
\subsection{Data}
The datasets used in this study encompass daytime and nighttime satellite imagery, along with five socioeconomic indicators from South Korea, Vietnam, Malawi, and Cambodia. The daytime images, sourced from WorldView-2 and GeoEye and captured between 2018 and 2022, include 406,754 images for South Korea, 967,317 for Vietnam, 336,361 for Malawi, and 512,976 for Cambodia. This study utilized 406,754 images for South Korea, 967,317 for VietNam, 336,361 for Malawi, and 512,976 for Cambodia. In total, 2,223,408 images were collected, each with a spatial resolution of about 2.4m and a size of 256x256 pixels. The nighttime images were sourced from the annual global Visible Infrared Imaging Radiometer Suite (VIIRS) nighttime lights provided by the Earth Observation Group (EOG). For this study, we utilized version 2.2 of VIIRS V2, the most recent version, which is continuously updated with data from recent years~\cite{elvidge2021annual}. This version offers comprehensive global coverage at a spatial resolution of approximately 500 meters as of 2022. The indicators—regional GDP, population, elderly population ratio, highly educated population ratio, and labor force participation rate—provided by each nation's government agency, were used to evaluate the model's performance. We collected data at a sub-national scale (district level for South Korea and Malawi, and province level for Vietnam and Cambodia), where data were accessible and provided by each nation's government agency.

\paragraph{Regional GDP (GRDP)} The regional GDP data for the year 2022 was collected from 229 districts in South Korea and 65 provinces in Vietnam. However, data at the subnational level was unavailable for other countries. This information was obtained from Statistics Korea and the Vietnam Law Library. \looseness=-1

\paragraph{Population (POP) \& Elderly population (ELP)} The population data for 2022, categorized into 15-year age intervals across all investigated countries, was sourced from the ESRI GeoEnrichment API. In this study, individuals aged 60 or older are classified as the elderly population.

\paragraph{Highly Educated Population Ratio (HER)} The highly educated population ratio represents the percentage of individuals who have achieved a bachelor’s degree relative to the entire population at all levels of educational attainment. The educational attainment data for 2022, covering each education level in South Korea, Malawi, and Vietnam, was collected from the ESRI GeoEnrichment API. However, equivalent data for Cambodia was not available from the API, prompting the use of data from the Demographic and Health Surveys program for 2021.

\paragraph{Labour force participation rate (LPR)} The labour force participation rate reflects the percentage of the working-age population (ages 15 to 64) who are either employed or actively seeking employment. The data for LPR covers the years 2019 to 2021 and was obtained from Statistics Korea (2021), the National Statistics Office of Malawi (2019), the General Statistics Office of Vietnam (2020), and the National Institute of Statistics of Cambodia (2019).

\newpage

\section{Ablation Study on Unsupervised Setting}
\label{sec:appendix_ablation}
Our model is composed of two main parts: module selection through the LLM and the subsequent data extraction from these modules for LLM inference.
Under an unsupervised setting, our ablation study removed or modified the following component one at a time:
(1) \textsf{PCA with all modules}: collecting numerical results from all modules in module list, except the address, to perform principal component analysis (PCA) to generate scores in the unsupervised setting; 
(2) \textsf{PCA with selected modules}: collecting numerical results from selected modules via our prompt, except the address, to perform PCA to generate scores in the unsupervised setting; 
(3) \textsf{GPT-4 Address only}: conducting LLM-based inference using only address information, without modules; 
(4) \textsf{GPT-4 Address with neighbor only}: performing LLM-based inference using our model with prompts used in GeoLLM~\cite{manvi2023geollm}; 
(5) \textsf{Without coarse-grained selection}: using only the fine-grained selection strategy (Section~\ref{sec:2.3}) to choose the same number of samples for in-context learning;
(6) \textsf{Without fine-grained selection}: using only the coarse-grained selection strategy;
(7) \textsf{Random selection}: randomly selecting in-context samples without any selection strategy.

Table~\ref{tab:appendix-ablation} displays the averaged absolute Pearson correlation across multiple socio-economic indicators for each country.
Consistently, any alterations or exclusions of components resulted in a decline in performance metrics on average. 
Similar to the findings in 5-shot settings (see Table~\ref{tab:ablation}), we observed that simply aggregating the module results for regression or omitting module results in LLM inference yields inferior performance compared to our full model.

\input{tables/appendix_ablation_study_unsupervised}

\newpage
\section{Module Selection Result}
\label{sec:appendix_module_selection}
Table~\ref{table:appendix_module_selection} reports module selections made by \model{} for each task reported in the main text. The rows indicate the modules utilized in our study. For a given task (column), one indicates that the module (row) was chosen and zero otherwise. We report the total number of modules used in each task at the bottom of the table. Utilization rate (last column) reports the share of tasks in which each module was selected by \model{}. 
\input{tables/appendix_discussion_module_selection}

\newpage
\section{Full Results of \model{}}
\label{sec:full_results}
The full evaluation results, including both Pearson ($\rho_p$) and Spearman ($\rho_s$) correlations, are presented from Table~\ref{table:appendix_full_unsup_perason_kv} to Table~\ref{table:appendix_full_few_pearson_mc}. 
We repeated the evaluations three times using random seeds. 
However, due to computational constraints, we conducted only one evaluation for South Korea, which has the largest number of regions. 
All numerical data are provided with three digits, including standard deviations, to clarify the statistical significance of our experiment. Some estimations from GPT-based models failed, reporting the same values across all regions. We excluded these cases when calculating averages, but if this occurred in all three experiments, we reported "N/A" in the table.

Tables~\ref{table:appendix_full_unsup_perason_kv}, \ref{table:appendix_full_unsup_perason_mc}, \ref{table:appendix_full_unsup_spearman_kv}, and \ref{table:appendix_full_unsup_spearman_mc} provide the full results for Table~\ref{table:main_unsupervised} in the main text, displaying the evaluation outcomes in the unsupervised setting.
Similarly, Tables~\ref{table:appendix_full_few_spearman_kv}, \ref{table:appendix_full_few_spearman_mc}, \ref{table:appendix_full_few_pearson_kv}, and \ref{table:appendix_full_few_pearson_mc} present the complete results for Table~\ref{table:main_fewshot_5} in the main text, showcasing the evaluation outcomes in the few-shot setting.

\input{tables/appendix_full_unsup_pearson_kv}
\input{tables/appendix_full_unsup_pearson_mc}
\input{tables/appendix_full_unsup_spearman_kv}
\input{tables/appendix_full_unsup_spearman_mc}

\input{tables/appendix_full_few_spearman_kv}
\input{tables/appendix_full_few_spearman_mc}
\input{tables/appendix_full_few_pearson_kv}
\input{tables/appendix_full_few_pearson_mc}

%% file: tables/appendix_ablation_study_unsupervised.tex
\begin{table}[h!]
\centering
\caption{Ablation study results. Averaged absolute Pearson correlations $|\rho_p|$ over the unsupervised setting for each country (i.e., South Korea - KOR, Malawi - MWI, Vietnam - VNM, Cambodia - KHM) across socio-economic indicators (i.e., POP, ELP, HER, LPR) after omitting or altering components of \model{}. \looseness=-1}
\label{tab:appendix-ablation}
\begin{tabular}{@{}lcccc|c@{}}
\toprule
Model & KOR & MWI & VNM & KHM & AVG \\ \midrule
Full model (\model{}) & 0.592 & 0.406 & 0.508 & 0.610 & \textbf{0.529} \\
(1) PCA with all modules & 0.502 & 0.237 & 0.209 & 0.568 & 0.379 \\
(2) PCA with selected modules & 0.559 & 0.455 & 0.369 & 0.650 & 0.508 \\
(3) GPT-4 Address only & 0.622 & 0.255 & 0.504 & 0.550 & 0.483 \\
(4) GPT-4 Address with neighbor only & 0.662 & 0.346 & 0.417 & 0.506 & 0.483 \\
(5) Without coarse-grained selection & 0.607 & 0.347 & 0.372 & 0.679 & 0.501 \\
(6) Without fine-grained selection & 0.551 & 0.342 & 0.358 & 0.568 & 0.455 \\
(7) Random selection & 0.586 & 0.372 & 0.343 & 0.536 & 0.459 \\ \bottomrule
\end{tabular}
\end{table}

%% file: tables/appendix_discussion_module_selection.tex
\begin{table}[h!]
\centering
\setlength{\tabcolsep}{2pt}
\caption{Module Selection by Task}
\label{table:appendix_module_selection}
\renewcommand{\arraystretch}{1.2}
\begin{adjustbox}{max width=\textwidth} 
\begin{threeparttable} \def\sym#1{\ifmmode^{#1}\else\(^{#1}\)\fi}
\begin{tabular}{@{}l|ccccc|cccc|ccccc|cccc|c@{}}
\toprule
\multirow{2}{*}{Module} & \multicolumn{5}{c|}{South Korea (KOR)} & \multicolumn{4}{c|}{Malawi (MWI)} & \multicolumn{5}{c|}{Vietnam (VNM)} & \multicolumn{4}{c|}{Cambodia (KHM)} & \multicolumn{1}{c}{Utilization} \\ \cline{2-19}
& GRDP & POP & ELP & HER & \multicolumn{1}{c|}{LPR} & POP & ELP & HER & LPR & GRDP & POP & ELP & HER & LPR & POP & ELP & HER & LPR &\multicolumn{1}{c}{rate} \\ \hline
Address & 1 & 1 & 1 & 1 & 1 & 1 & 1 & 1 & 1 & 1 & 1 & 1 & 1 & 1 & 1 & 1 & 1 & 1 & 1.000 \\
Area & 1 & 1 & 1 & 1 & 1 & 1 & 1 & 1 & 1 & 1 & 1 & 1 & 1 & 1 & 1 & 1 & 1 & 1 & 1.000 \\
Nightlight & 1 & 1 & 1 & 1 & 1 & 1 & 1 & 1 & 1 & 1 & 1 & 1 & 1 & 1 & 1 & 1 & 1 & 1 & 1.000 \\
Landcover &  &  &  &  &  &  &  &  &  &  & &  &  &  &  &  &  &  &  \\
\quad Agriculture & 0 & 1 & 0 & 0 & 1 & 1 & 1 & 0 & 1 & 1 & 0 & 0 & 0 & 1 & 1 & 0 & 0 & 1 & 0.500 \\
\quad Building & 1 & 1 & 1 & 1 & 1 & 1 & 1 & 1 & 1 & 1 & 1 & 1 & 1 & 1 & 1 & 1 & 1 & 1 & 1.000 \\
\quad Development & 1 & 1 & 1 & 1 & 1 & 1 & 1 & 1 & 1 & 1 & 1 & 1 & 1 & 1 & 1 & 1 & 1 & 1 &  1.000 \\
\quad Rangeland & 0 & 1 & 1 & 0 & 0 & 1 & 1 & 0 & 0 & 0 & 0 & 1 & 0 & 0 & 1 & 1 & 0 & 0 & 0.389 \\
\quad Road & 0 & 1 & 0 & 0 & 0 & 0 & 0 & 0 & 0 & 1 & 1 & 0 & 0 & 0 & 0 & 0 & 0 & 0 & 0.167 \\
\quad Water & 0 & 0 & 0 & 0 & 0 & 0 & 0 & 0 & 0 & 0 & 0& 0 & 0 & 0 &  0& 0 & 0 & 0 & 0.000 \\
Distance &  &  &  &  &  &  &  &  &  &  & &  &  &  &  &  &  &  &  \\
\quad Airport & 1 & 0 & 0 & 0 & 1 & 0 & 0 & 0 & 0 & 1 & 0& 0 & 1 & 1 & 0 & 0 & 1 & 1 & 0.389 \\
\quad Port & 0 & 0 & 0 & 0 & 1 & 0 & 0 & 0 & 0 & 1 & 0& 0 & 1 & 1 & 0 & 0 & 1 & 1 & 0.333 \\
Neighbor & & &  &  &  &  &  &  &  &  & &  &  &  &  &  &  &  &  \\
\quad Area & 0 & 0 & 1 & 0 & 0 & 0 & 0 & 0 & 0 & 0 & 0 & 1 & 0 & 0 & 0 & 0 & 0 & 0 & 0.111 \\
\quad Nightlight & 1 & 1 & 1 & 1 & 0 & 1 & 1 & 1 & 0 & 1 & 1 & 1 & 1 & 0 & 1 & 1 & 1 & 1 & 0.833 \\
\quad Agriculture & 0 & 0 & 0 & 0 & 1 & 1 & 0 & 0 & 1 & 1 & 0 & 0 & 0 & 1 & 1 & 0 & 0 & 1 & 0.389 \\
\quad Building & 1 & 1 & 1 & 1 & 0 & 1 & 1 & 1 & 0 & 0 & 1 & 1 & 1 & 0 & 1 & 1 & 0 & 0 & 0.667 \\
\quad Development & 1 & 1 & 1 & 1 & 1 & 1 & 1 & 1 & 1 & 1 & 1 & 1 & 1 & 1 & 1 & 1 & 1 & 1 &  1.000 \\
\quad Rangeland & 0 & 0 & 0 & 0 & 0 & 1 & 0 & 0 & 0 & 0 & 0 & 0 & 0 & 0 & 1 & 0 & 0 & 0 & 0.111 \\
\quad Road & 0 & 1 & 0 & 0 & 0 & 0 & 0 & 0 & 0 & 0 & 0 & 0 & 0 & 0 & 0 & 0 & 0 & 0 & 0.056 \\
\quad Water & 0 & 0 & 0 & 0 & 0 & 0 & 0 & 0 & 0 & 0 & 0& 0 & 0 & 0 & 0 & 0 & 0 & 0 & 0.000 \\
\midrule
Total modules & 9 & 12 & 10 & 8 & 10 & 12 & 10 & 8 & 8 & 12 & 9 & 10 & 10 & 10 & 12 & 9 & 9 & 11 &  \\
\bottomrule
\end{tabular}
\end{threeparttable}
\end{adjustbox}
\end{table}

%% file: tables/appendix_full_unsup_pearson_kv.tex
\begin{table}[ht!]
\centering
\setlength{\tabcolsep}{2pt}
\caption{Performance evaluation results with Pearson correlation |$\rho_p$| in unsupervised setting for South Korea and Viet Nam. Regional GDP (GRDP) data is accessible for these countries.}
\label{table:appendix_full_unsup_perason_kv}
\renewcommand{\arraystretch}{1.4}
\resizebox{1\textwidth}{!}{
\begin{tabular}{l|ccccc|ccccc}
\toprule
\multirow{2}{*}{Method} & \multicolumn{5}{c|}{South Korea} & \multicolumn{5}{c}{Vietnam}\\ \cline{2-11}
 & GRDP & POP & ELP & HER & LPR & GRDP & POP & ELP & HER & LPR\\ \hline
Nightlight & 0.672 & 0.696 & 0.632 & 0.535 & 0.548 & 0.251 & 0.254 & 0.225 & 0.584 & 0.009\\
SiScore & 0.347$\pm$0.007 & 0.458$\pm$0.012 & 0.49$\pm$0.01 & 0.643$\pm$0.004 & 0.7$\pm$0.017 & 0.388$\pm$0.008 & 0.366$\pm$0.007 & 0.449$\pm$0.01 & 0.078$\pm$0.012 & 0.138$\pm$0.007\\
UrbanScore & 0.263$\pm$0.04 & 0.385$\pm$0.043 & 0.424$\pm$0.041 & 0.606$\pm$0.027 & 0.658$\pm$0.049 & 0.475$\pm$0.107 & 0.419$\pm$0.086 & 0.455$\pm$0.079 & 0.426$\pm$0.232 & 0.092$\pm$0.016\\
GPT-4-Wiki & 0.433 & 0.543 & 0.314 & 0.562 & 0.278 & 0.353$\pm$0.01 & 0.67$\pm$0.248 & 0.224$\pm$0.02 & 0.499$\pm$0.003 & 0.035$\pm$0.052\\
\model{} & 0.535 & 0.625 & 0.471 & 0.653 & 0.62 & 0.525$\pm$0.166 & 0.817$\pm$0.01 & 0.634$\pm$0.301 & 0.221$\pm$0.009 & 0.361$\pm$0.045\\
\bottomrule
\end{tabular}
}
\end{table}

%% file: tables/appendix_full_unsup_pearson_mc.tex
\begin{table}[ht!]
\centering
\setlength{\tabcolsep}{2pt}
\caption{Performance evaluation results with Pearson correlation |$\rho_p$| in unsupervised setting for Malawi and Cambodia.}
\label{table:appendix_full_unsup_perason_mc}
\renewcommand{\arraystretch}{1.4}
\resizebox{1\textwidth}{!}{
\begin{tabular}{l|cccc|cccc}
\toprule
\multirow{2}{*}{Method} & \multicolumn{4}{c|}{Malawi} & \multicolumn{4}{c}{Cambodia}\\ \cline{2-9}
 & POP & ELP & HER & LPR & POP & ELP & HER & LPR\\ \hline
Nightlight & 0.428 & 0.085 & 0.86 & 0.192 & 0.761 & 0.644 & 0.829 & 0.510\\
SiScore & 0.029$\pm$0.009 & 0.178$\pm$0.024 & 0.795$\pm$0.041 & 0.114$\pm$0.029 & 0.746$\pm$0.011 & 0.742$\pm$0.017 & 0.736$\pm$0.033 & 0.356$\pm$0.019\\
UrbanScore & 0.081$\pm$0.069 & 0.272$\pm$0.037 & 0.754$\pm$0.268 & 0.214$\pm$0.057 & 0.639$\pm$0.109 & 0.6$\pm$0.117 & 0.762$\pm$0.07 & 0.43$\pm$0.084\\
GPT-4-Wiki & 0.327$\pm$0.045 & 0.286$\pm$0.006 & 0.653$\pm$0.054 & 0.282$\pm$0.082 & 0.605$\pm$0.129 & 0.29$\pm$0.009 & 0.25$\pm$0.099 & 0.57$\pm$0.041\\
\model{} & 0.374$\pm$0.061 & 0.163$\pm$0.02 & 0.848$\pm$0.045 & 0.24$\pm$0.038 & 0.784$\pm$0.08 & 0.702$\pm$0.05 & 0.608$\pm$0.006 & 0.345$\pm$0.027\\
\bottomrule
\end{tabular}
}
\end{table}

%% file: tables/appendix_full_unsup_spearman_kv.tex
\begin{table}[ht!]
\centering
\setlength{\tabcolsep}{2pt}
\caption{Performance evaluation results with Spearman correlation |$\rho_s$| in unsupervised setting for South Korea and Viet Nam. Regional GDP (GRDP) data is accessible for these countries.}
\label{table:appendix_full_unsup_spearman_kv}
\renewcommand{\arraystretch}{1.4}
\resizebox{1\textwidth}{!}{
\begin{tabular}{l|ccccc|ccccc}
\toprule
\multirow{2}{*}{Method} & \multicolumn{5}{c|}{South Korea} & \multicolumn{5}{c}{Vietnam}\\ \cline{2-11}
 & GRDP & POP & ELP & HER & LPR & GRDP & POP & ELP & HER & LPR\\ \hline
Nightlight & 0.761 & 0.691 & 0.65 & 0.734 & 0.758 & 0.870 & 0.767 & 0.729 & 0.027 & 0.164\\
SiScore & 0.684$\pm$0.002 & 0.742$\pm$0.003 & 0.712$\pm$0.001 & 0.746$\pm$0.006 & 0.794$\pm$0.004 & 0.569$\pm$0.007 & 0.513$\pm$0.005 & 0.625$\pm$0.007 & 0.235$\pm$0.011 & 0.037$\pm$0.005\\
UrbanScore & 0.524$\pm$0.115 & 0.603$\pm$0.106 & 0.58$\pm$0.099 & 0.685$\pm$0.072 & 0.717$\pm$0.09 & 0.582$\pm$0.082 & 0.427$\pm$0.094 & 0.546$\pm$0.094 & 0.296$\pm$0.107 & 0.048$\pm$0.036\\
GPT-4-Wiki & 0.607 & 0.508 & 0.336 & 0.611 & 0.289 & 0.413$\pm$0.046 & 0.168$\pm$0.012 & 0.161$\pm$0.059 & 0.207$\pm$0.01 & 0.0$\pm$0.025\\
\model{} & 0.8 & 0.765 & 0.692 & 0.703 & 0.649 & 0.822$\pm$0.021 & 0.762$\pm$0.005 & 0.736$\pm$0.056 & 0.07$\pm$0.017 & 0.314$\pm$0.067\\
\bottomrule
\end{tabular}
}
\end{table}

%% file: tables/appendix_full_unsup_spearman_mc.tex
\begin{table}[ht!]
\centering
\setlength{\tabcolsep}{2pt}
\caption{Performance evaluation results with Spearman correlation |$\rho_s$| in unsupervised setting for Malawi and Cambodia.}
\label{table:appendix_full_unsup_spearman_mc}
\renewcommand{\arraystretch}{1.4}
\resizebox{1\textwidth}{!}{
\begin{tabular}{l|cccc|cccc}
\toprule
\multirow{2}{*}{Method} & \multicolumn{4}{c|}{Malawi} & \multicolumn{4}{c}{Cambodia}\\ \cline{2-9}
 & POP & ELP & HER & LPR & POP & ELP & HER & LPR\\ \hline
Nightlight & 0.657 & 0.492 & 0.331 & 0.046 & 0.753 & 0.708 & 0.525 & 0.404\\
SiScore & 0.196$\pm$0.019 & 0.085$\pm$0.015 & 0.036$\pm$0.004 & 0.011$\pm$0.014 & 0.646$\pm$0.012 & 0.675$\pm$0.020 & 0.434$\pm$0.029 & 0.298$\pm$0.016\\
UrbanScore & 0.184$\pm$0.098 & 0.268$\pm$0.091 & 0.33$\pm$0.128 & 0.209$\pm$0.113 & 0.403$\pm$0.128 & 0.442$\pm$0.118 & 0.526$\pm$0.058 & 0.299$\pm$0.112\\
GPT-4-Wiki & 0.182$\pm$0.017 & 0.286$\pm$0.050 & 0.309$\pm$0.064 & 0.280$\pm$0.066 & 0.314$\pm$0.104 & 0.231$\pm$0.012 & 0.533$\pm$0.091 & 0.533$\pm$0.035\\
\model{} & 0.505$\pm$0.122 & 0.311$\pm$0.008 & 0.229$\pm$0.12 & 0.26$\pm$0.030 & 0.692$\pm$0.104 & 0.592$\pm$0.087 & 0.350$\pm$0.039 & 0.334$\pm$0.002\\
\bottomrule
\end{tabular}
}
\end{table}

%% file: tables/appendix_full_few_spearman_kv.tex
\begin{table}[ht!]
\centering
\setlength{\tabcolsep}{2pt}
\caption{Full evaluation results with Spearman correlation $\rho_s$ in 5-shots setting for South Korea and Viet Nam, with 3 digits. Regional GDP (GRDP) data is accessible for these countries. }
\label{table:appendix_full_few_spearman_kv}
\renewcommand{\arraystretch}{1.4}
\resizebox{1\textwidth}{!}{
\begin{tabular}{l|ccccc|ccccc}
\toprule
\multirow{2}{*}{Method} & \multicolumn{5}{c|}{South Korea} & \multicolumn{5}{c}{Vietnam}\\ \cline{2-11}
 & GRDP & POP & ELP & HER & LPR & GRDP & POP & ELP & HER & LPR\\ \hline
Nightlight & 0.251$\pm$0.888 & 0.697$\pm$0.006 & 0.656$\pm$0.006 & 0.733$\pm$0.006 & 0.758$\pm$0.002 & 0.871$\pm$0.010 & 0.773$\pm$0.028 & 0.717$\pm$0.010 & -0.039$\pm$0.050 & -0.077$\pm$0.195\\
SimpleCNN & 0.311$\pm$0.203 & 0.143$\pm$0.491 & 0.277$\pm$0.164 & 0.525$\pm$0.217 & 0.256$\pm$0.460 & 0.096$\pm$0.243 & 0.271$\pm$0.229 & 0.433$\pm$0.283 & 0.197$\pm$0.134 & 0.372$\pm$0.074\\
READ & 0.562$\pm$0.080 & 0.318$\pm$0.484 & 0.288$\pm$0.490 & 0.341$\pm$0.330 & 0.307$\pm$0.310 & 0.611$\pm$0.072 & 0.342$\pm$0.078 & 0.212$\pm$0.412 & -0.176$\pm$0.114 & 0.138$\pm$0.215\\
Tile2Vec & 0.271$\pm$0.206 & 0.306$\pm$0.338 & 0.189$\pm$0.368 & 0.578$\pm$0.150 & 0.345$\pm$0.147 & 0.126$\pm$0.258 & 0.383$\pm$0.173 & 0.194$\pm$0.482 & 0.149$\pm$0.120 & 0.368$\pm$0.078\\
SimCLR & 0.144$\pm$0.092 & 0.223$\pm$0.122 & 0.207$\pm$0.071 & 0.464$\pm$0.086 & 0.080$\pm$0.194 & 0.303$\pm$0.178 & 0.228$\pm$0.146 & 0.344$\pm$0.112 & 0.048$\pm$0.191 & -0.164$\pm$0.128\\
GeoLLM & 0.205 & 0.755 & 0.465 & 0.635 & -0.146 & 0.752$\pm$0.067 & 0.785$\pm$0.045 & 0.471$\pm$0.270 & 0.428$\pm$0.273 & -0.012\\
GPT-4-Wiki & 0.478 & 0.534 & 0.473 & 0.656 & -0.162 & 0.490$\pm$0.112 & 0.240$\pm$0.150 & 0.151$\pm$0.032 & 0.221$\pm$0.055 & 0.038$\pm$0.028\\
\model{} & 0.812 & 0.856 & 0.805 & 0.774 & 0.683 & 0.888$\pm$0.015 & 0.979$\pm$0.005 & 0.834$\pm$0.049 & 0.427$\pm$0.059 & 0.238$\pm$0.100\\
\bottomrule
\end{tabular}
}
\end{table}

%% file: tables/appendix_full_few_spearman_mc.tex
\begin{table}[ht!]
\centering
\setlength{\tabcolsep}{2pt}
\caption{Full evaluation results with Spearman correlation $\rho_s$ in 5-shots setting for Malawi and Cambodia, with 3 digits.}
\label{table:appendix_full_few_spearman_mc}
\renewcommand{\arraystretch}{1.4}
\resizebox{1\textwidth}{!}{
\begin{tabular}{l|cccc|cccc}
\toprule
\multirow{2}{*}{Method} & \multicolumn{4}{c|}{Malawi} & \multicolumn{4}{c}{Cambodia}\\ \cline{2-9}
 & POP & ELP & HER & LPR & POP & ELP & HER & LPR\\ \hline
Nightlight & 0.216$\pm$0.810 & 0.145$\pm$0.603 & 0.045$\pm$0.437 & -0.006$\pm$0.049 & 0.728$\pm$0.085 & 0.674$\pm$0.074 & 0.514$\pm$0.088 & 0.409$\pm$0.060\\
SimpleCNN & 0.387$\pm$0.058 & 0.514$\pm$0.043 & 0.121$\pm$0.154 & 0.208$\pm$0.407 & 0.300$\pm$0.443 & 0.285$\pm$0.481 & -0.218$\pm$0.244 & 0.237$\pm$0.078\\
READ & 0.047$\pm$0.079 & 0.054$\pm$0.116 & 0.125$\pm$0.103 & 0.244$\pm$0.064 & 0.252$\pm$0.119 & 0.298$\pm$0.016 & 0.137$\pm$0.116 & -0.047$\pm$0.151\\
Tile2Vec & 0.356$\pm$0.037 & 0.379$\pm$0.118 & 0.212$\pm$0.166 & 0.299$\pm$0.320 & 0.100$\pm$0.661 & 0.063$\pm$0.280 & 0.245$\pm$0.088 & 0.290$\pm$0.106\\
SimCLR & 0.142$\pm$0.335 & 0.123$\pm$0.202 & 0.238$\pm$0.138 & 0.093$\pm$0.202 & 0.410$\pm$0.171 & 0.437$\pm$0.132 & 0.005$\pm$0.198 & 0.180$\pm$0.060\\
GeoLLM & 0.271$\pm$0.107 & 0.506$\pm$0.220 & 0.292$\pm$0.000 & N/A & 0.384$\pm$0.031 & -0.110$\pm$0.558 & 0.439$\pm$0.000 & 0.088$\pm$0.130\\
GPT-4-Wiki & 0.506$\pm$0.171 & 0.391$\pm$0.028 & 0.360$\pm$0.143 & 0.140$\pm$0.035 & 0.290$\pm$0.254 & 0.494$\pm$0.076 & 0.505$\pm$0.054 & -0.213$\pm$0.141\\
\model{} & 0.766$\pm$0.224 & 0.507$\pm$0.070 & 0.271$\pm$0.068 & -0.275$\pm$0.14 & 0.643$\pm$0.174 & 0.826$\pm$0.031 & 0.467$\pm$0.044 & 0.367$\pm$0.166\\
\bottomrule
\end{tabular}
}
\end{table}

%% file: tables/appendix_full_few_pearson_kv.tex
\begin{table}[ht!]
\centering
\setlength{\tabcolsep}{2pt}
\caption{Full evaluation results with Pearson correlation $\rho_p$ in 5-shots setting for South Korea and Viet Nam, with 3 digits. Regional GDP (GRDP) data is accessible for these countries.}
\label{table:appendix_full_few_pearson_kv}
\renewcommand{\arraystretch}{1.4}
\resizebox{1\textwidth}{!}{
\begin{tabular}{l|ccccc|ccccc}
\toprule
\multirow{2}{*}{Method} & \multicolumn{5}{c|}{South Korea} & \multicolumn{5}{c}{Vietnam}\\ \cline{2-11}
 & GRDP & POP & ELP & HER & LPR & GRDP & POP & ELP & HER & LPR\\ \hline
Nightlight & 0.125$\pm$0.696 & 0.545$\pm$0.262 & 0.477$\pm$0.258 & 0.534$\pm$0.005 & 0.548$\pm$0.003 & 0.017$\pm$0.086 & 0.050$\pm$0.134 & 0.021$\pm$0.108 & -0.217$\pm$0.686 & -0.009$\pm$0.018\\
SimpleCNN & 0.371$\pm$0.228 & 0.236$\pm$0.411 & 0.304$\pm$0.305 & 0.486$\pm$0.263 & 0.241$\pm$0.380 & -0.022$\pm$0.045 & 0.186$\pm$0.139 & 0.406$\pm$0.201 & 0.028$\pm$0.181 & 0.405$\pm$0.113\\
READ & 0.513$\pm$0.092 & 0.216$\pm$0.324 & 0.237$\pm$0.324 & 0.253$\pm$0.291 & 0.253$\pm$0.289 & 0.398$\pm$0.136 & 0.264$\pm$0.023 & 0.156$\pm$0.104 & -0.281$\pm$0.139 & 0.056$\pm$0.226\\
Tile2Vec & 0.444$\pm$0.106 & 0.424$\pm$0.291 & 0.294$\pm$0.345 & 0.557$\pm$0.159 & 0.288$\pm$0.194 & -0.020$\pm$0.096 & 0.137$\pm$0.213 & 0.243$\pm$0.232 & 0.052$\pm$0.134 & 0.516$\pm$0.096\\
SimCLR & 0.183$\pm$0.078 & 0.227$\pm$0.028 & 0.242$\pm$0.030 & 0.333$\pm$0.141 & 0.023$\pm$0.165 & 0.211$\pm$0.297 & 0.051$\pm$0.176 & 0.162$\pm$0.060 & -0.003$\pm$0.222 & -0.163$\pm$0.228\\
GeoLLM & -0.016 & 0.634 & 0.010 & -0.020 & -0.104 & 0.228$\pm$0.405 & 0.843$\pm$0.083 & 0.795$\pm$0.083 & 0.560$\pm$0.467 & 0.013$\pm$0.000\\
GPT-4-Wiki & 0.431 & 0.589 & 0.414 & 0.561 & -0.261 & 0.435$\pm$0.106 & 0.827$\pm$0.014 & 0.805$\pm$0.024 & 0.471$\pm$0.129 & 0.114$\pm$0.074\\
\model{} & 0.792 & 0.671 & 0.752 & 0.708 & 0.686 & 0.941$\pm$0.010 & 0.998$\pm$0.000 & 0.901$\pm$0.046 & 0.646$\pm$0.148 & 0.325$\pm$0.075\\
\bottomrule
\end{tabular}
}
\end{table}

%% file: tables/appendix_full_few_pearson_mc.tex
\begin{table}[ht!]
\centering
\setlength{\tabcolsep}{2pt}
\caption{Full evaluation results with Pearson correlation $\rho_p$ in 5-shots setting for Malawi and Cambodia, with 3 digits.}
\label{table:appendix_full_few_pearson_mc}
\renewcommand{\arraystretch}{1.4}
\resizebox{1\textwidth}{!}{
\begin{tabular}{l|cccc|cccc}
\toprule
\multirow{2}{*}{Method} & \multicolumn{4}{c|}{Malawi} & \multicolumn{4}{c}{Cambodia}\\ \cline{2-9}
 & POP & ELP & HER & LPR & POP & ELP & HER & LPR\\ \hline
Nightlight & -0.018$\pm$0.534 & -0.137$\pm$0.179 & 0.305$\pm$1.007 & -0.081$\pm$0.214 & 0.676$\pm$0.101 & 0.587$\pm$0.035 & 0.705$\pm$0.236 & 0.441$\pm$0.209\\
SimpleCNN & 0.370$\pm$0.102 & 0.492$\pm$0.109 & 0.050$\pm$0.128 & 0.229$\pm$0.390 & 0.038$\pm$0.316 & 0.161$\pm$0.473 & -0.296$\pm$0.114 & 0.157$\pm$0.145\\
READ & -0.061$\pm$0.099 & -0.044$\pm$0.117 & 0.485$\pm$0.077 & 0.240$\pm$0.109 & 0.258$\pm$0.114 & 0.282$\pm$0.196 & 0.173$\pm$0.163 & -0.058$\pm$0.076\\
Tile2Vec & 0.257$\pm$0.158 & 0.059$\pm$0.330 & 0.227$\pm$0.466 & 0.310$\pm$0.230 & -0.009$\pm$0.267 & -0.154$\pm$0.269 & 0.220$\pm$0.063 & 0.221$\pm$0.100\\
SimCLR & 0.202$\pm$0.261 & 0.128$\pm$0.262 & 0.321$\pm$0.209 & 0.159$\pm$0.121 & 0.418$\pm$0.110 & 0.395$\pm$0.265 & 0.144$\pm$0.329 & 0.142$\pm$0.035\\
GeoLLM & 0.401$\pm$0.040 & 0.494$\pm$0.149 & -0.004$\pm$0.000 & N/A & 0.330$\pm$0.070 & 0.118$\pm$0.002 & 0.480$\pm$0.000 & 0.111$\pm$0.105\\
GPT-4-Wiki & 0.520$\pm$0.171 & 0.577$\pm$0.006 & 0.699$\pm$0.081 & 0.147$\pm$0.065 & 0.553$\pm$0.241 & 0.636$\pm$0.085 & 0.539$\pm$0.071 & -0.215$\pm$0.120\\
\model{} & 0.753$\pm$0.189 & 0.513$\pm$0.116 & 0.938$\pm$0.053 & -0.261$\pm$0.093 & 0.734$\pm$0.116 & 0.791$\pm$0.128 & 0.828$\pm$0.015 & 0.603$\pm$0.019\\
\bottomrule
\end{tabular}
}
\end{table}